\documentclass[12pt,a4paper]{article}
\usepackage[dvips]{graphicx}

\makeatletter
\newif\if@preliminary
\@preliminaryfalse
\def\preliminary{\@preliminarytrue}
\def\preprintno#1{\def\@preprintno{#1}}
\def\address#1{\def\@address{#1}}

\def\abstract#1{\def\@abstract{#1}}
\renewcommand\abstractname{ABSTRACT}
\newlength\preprintnoskip
\newlength\abstractwidth
\@titlepagetrue
\renewcommand\maketitle{\begin{titlepage}%
  \let\footnotesize\small
  \hfill\parbox{\preprintnoskip}{%
  \begin{flushright}\@preprintno\end{flushright}}\hspace*{1cm}
  \vskip 60\p@
  \begin{center}%
    {\Large\bf\boldmath \@title \par}\vskip 1cm%
    {\sc\@author \par}\vskip 3mm%
    {\@address \par}%
    \if@preliminary
      \vskip 2cm {\large\sf PRELIMINARY DRAFT \par \@date}%
    \fi
  \end{center}\par
  \@thanks
  \vfill
  \begin{center}%
    \parbox{\abstractwidth}{\centerline{\abstractname}%
    \vskip 3mm%
    \@abstract}
  \end{center}
  \end{titlepage}%
  \setcounter{footnote}{0}%
  \let\thanks\relax\let\maketitle\relax
  \gdef\@thanks{}\gdef\@author{}\gdef\@address{}%
  \gdef\@title{}\gdef\@abstract{}\gdef\@preprintno{}
}%
\def\shortletter{%
  \setcounter{secnumdepth}{5}
  \def\paragraph{%
    \@startsection{paragraph}{4}{\parindent}%
      {3.25ex \@plus1ex \@minus.2ex}{-.5em}%
      {\reset@font\normalsize\bfseries}}%
  \renewcommand\theparagraph{\arabic{paragraph}.\hskip-.5em}
  \def\subparagraph{%
    \@startsection{subparagraph}{5}{\parindent}%
      {3.25ex \@plus1ex \@minus.2ex}{-.5em}%
      {\reset@font\normalsize\bfseries}}%
  \renewcommand\thesubparagraph{(\alph{subparagraph})\hskip-.5em}
}
\def\thesection{\arabic{section}.}

\def\appendix{\setcounter{section}{0}
 \def\thesection{Appendix \Alph{section}:}
 \def\theequation{\Alph{section}.\arabic{equation}}}
\def\@citex[#1]#2{\if@filesw\immediate\write\@auxout{\string\citation{#2}}\fi
  \def\@citea{}\@cite{\@for\@citeb:=#2\do
    {\@citea\def\@citea{,\penalty\@m}\@ifundefined
       {b@\@citeb}{{\bf ?}\@warning
       {Citation `\@citeb' on page \thepage \space undefined}}%
\hbox{\csname b@\@citeb\endcsname}}}{#1}}
\def\citerange{\@ifnextchar [{\@tempswatrue\@citexr}{\@tempswafalse\@citexr[]}}
\def\@citexr[#1]#2{\if@filesw\immediate\write\@auxout{\string\citation{#2}}\fi
  \def\@citea{}\@cite{\@for\@citeb:=#2\do
    {\@citea\def\@citea{--\penalty\@m}\@ifundefined
       {b@\@citeb}{{\bf ?}\@warning
       {Citation `\@citeb' on page \thepage \space undefined}}%
\hbox{\csname b@\@citeb\endcsname}}}{#1}}
\long\def\@makecaption#1#2{%
  \vskip\abovecaptionskip
  \sbox\@tempboxa{#1: \emph{#2}}%
  \ifdim \wd\@tempboxa >\hsize
    #1: \emph{#2}\par
  \else
    \hbox to\hsize{\hfil\box\@tempboxa\hfil}%
  \fi
  \vskip\belowcaptionskip}
\def\fmslash{\@ifnextchar[{\fmsl@sh}{\fmsl@sh[0mu]}}
\def\fmsl@sh[#1]#2{%
  \mathchoice
    {\@fmsl@sh\displaystyle{#1}{#2}}%
    {\@fmsl@sh\textstyle{#1}{#2}}%
    {\@fmsl@sh\scriptstyle{#1}{#2}}%
    {\@fmsl@sh\scriptscriptstyle{#1}{#2}}}
\def\@fmsl@sh#1#2#3{\m@th\ooalign{$\hfil#1\mkern#2/\hfil$\crcr$#1#3$}}
\makeatother

\def\fmfL(#1,#2,#3)#4{\put(#1,#2){\makebox(0,0)[#3]{#4}}}

\begin{document}
\shortletter        
\baselineskip20pt   
\preprintno{DESY 95--216\\hep-ph/9512355\\December 1995}
\title{%
 HIGGS-STRAHLUNG AND $WW$ FUSION\\
 IN $e^+e^-$ COLLISIONS
}
\author{%
 W.~Kilian,
 M.~Kr\"amer,
 and P.M.~Zerwas
}
\address{%
 Deutsches Elektronen-Synchrotron DESY\\
 D-22603 Hamburg/FRG
}
\abstract{%
Higgs-strahlung $e^+e^-\to ZH$ and $WW$ fusion
$e^+e^-\to\bar\nu_e\nu_e H$ are the most important mechanisms for the
production of Higgs bosons in $e^+e^-$ collisions at LEP2 and future
$e^+e^-$ linear colliders.  We have calculated the cross sections and
energy/angular distributions of the Higgs boson for these production
mechanisms.  When the $Z$ boson decays into (electron-)neutrinos, the
two production amplitudes interfere.  In the cross-over region between
the two mechanisms the interference term is positive and of the same
size as the individual cross sections, thus enhancing the production
rate.
}
\maketitle

\paragraph{}
The analysis of the mechanism which breaks the electroweak gauge
symmetry $SU(2)_{\rm L}\times U(1)_{\rm Y}$ down to $U(1)_{\rm EM}$,
is one of the key problems in particle physics.  If the gauge fields
involved remain weakly interacting up to high energies -- a
prerequisite for the (perturbative) renormalization of
$\sin^2\theta_W$ from the symmetry value $3/8$ of grand-unified
theories down to a value near $0.2$ at low energies -- fundamental
scalar Higgs bosons~\cite{Higgs} must exist which damp the rise of the
scattering amplitudes of massive gauge particles at high energies.  In
the Standard Model (SM) an isoscalar doublet field is introduced to
accomodate the electroweak data, leading to the prediction of a single
Higgs boson.  Supersymmetric extensions of the Standard Model expand
the scalar sector to a spectrum of Higgs particles~\cite{SUSY}.  The
Higgs particles have been searched for, unsuccessfully so far, at
LEP1, setting a lower limit on the SM Higgs mass of $m_H>65.2$
GeV~\cite{Limit}.  The search for these particles and, if found, the
exploration of their profile, will continue at LEP2~\cite{LEP2}, the
LHC~\cite{LHC}, and future $e^+e^-$ linear colliders~\cite{FLC}.

\begin{figure}[bht]
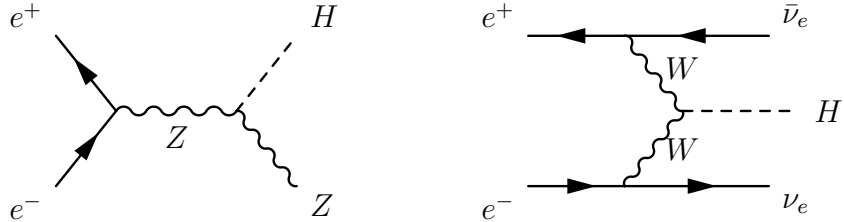

\unitlength 1mm
\begin{center}
\begin{picture}(40,20)
  \put(0,0){\includegraphics{fmgraphs.1}}
\fmfL(19.99997,7.89122,t){$Z$}%
\fmfL(2.21202,-1.11765,rt){$e^-$}%
\fmfL(2.21202,21.11765,rb){$e^+$}%
\fmfL(37.78798,-1.11765,lt){$Z$}%
\fmfL(37.78798,21.11765,lb){$H$}%
\end{picture}
\hspace{2cm}
\begin{picture}(40,20)
  \put(0,0){\includegraphics{fmgraphs.2}}
\fmfL(24.53337,6.29999,t){$W$}%
\fmfL(2.21202,-1.11765,rt){$e^-$}%
\fmfL(2.21202,21.11765,rb){$e^+$}%
\fmfL(37.78798,-1.11765,lt){$\nu _e$}%
\fmfL(42.10876,10,l){$H$}%
\fmfL(37.78798,21.11765,lb){$\bar \nu _e$}%
\fmfL(24.53337,14.20003,b){$W$}%
\end{picture}
\end{center}
\caption{Higgs-strahlung and $WW$ fusion of (CP--even) Higgs bosons in
$e^+e^-$ collisions.}
\label{fig:graphs}
\end{figure}
In this note we will focus on the production of scalar Higgs bosons
in $e^+e^-$ collisions.  The main production mechanisms for these
particles are Higgs-strahlung~\cite{hst} and $WW$
fusion~\cite{wwfi,wwf,AMP} [supplemented in supersymmetric theories
by associated scalar/pseudoscalar Higgs pair production].  In
particular, we will present a comprehensive analysis of the interplay
between the production mechanisms%
\footnote{
  We will concentrate
  first on the Standard Model (SM); the extension to the Minimal
  Supersymmetric Standard Model (MSSM) is trivial as will be
  demonstrated in the last section of this note.}
(Fig.\ref{fig:graphs})
\begin{equation}
  \begin{array}{lll}
  \mbox{Higgs-strahlung} \hskip-.75em &: & e^+e^-\to ZH \to \bar\nu\nu H
  \\[2mm]
  \mbox{$WW$ fusion}&: & e^+e^- \to \bar\nu_e\nu_e H
  \end{array}
\end{equation}
if the $Z$ bosons decay into neutrinos.  For $\bar\nu_e\nu_e$ decays
of the $Z$ bosons, the two production amplitudes interfere.  It turns
out that the interference term is positive and of the same size as the
individual cross sections in the cross-over region between the two
mechanisms.  Thus, the interference term adds to the rate at LEP2
where the fusion mechanism will be exploited to drive the discovery
range of Higgs particles to the ultimate experimental
limit~\cite{LEP2,ULim}.  The interference effect had been noticed
earlier~\cite{wwfi,boos}; however, we improve on these calculations by
deriving an analytic result for the energy and polar angular
distribution of the Higgs particle $(E_H,\theta)$ in the final state
of $e^+e^- \to H + \mbox{neutrinos}$.  This representation can
comfortably serve as input for Monte Carlo generators like
PYTHIA/JETSET~\cite{PYTHIA} and HZHA~\cite{HZHA} which include the
leading QED bremsstrahlung corrections and the important background
processes.

\paragraph{}
 The cross section for the Higgs-strahlung process can be written in
the following compact form:
\begin{equation}\label{eq:HZ}
  \sigma(e^{+}e^{-}\to ZH) =
  \frac{G_{F}^{2}m_{Z}^{4}}{96 \pi s}\,\left( v_{e}^{2}+a_{e}^{2}\right)
  \,\lambda^{\frac{1}{2}}\,
  \frac{\lambda + 12m_{Z}^{2}/s}{(1-m_{Z}^{2}/s)^{2}}
\end{equation}
where $\sqrt{s}$ is the center-of-mass energy, and $a_e=-1$,
$v_e=-1+4\sin^2\theta_W$ are the $Z$ charges of the electron; $\lambda
= \left(1-(m_H+m_Z)^2/s\right)\left(1-(m_H-m_Z)^2/s\right)$ is the usual
two-particle phase space function.  So long as the non-zero width of
the $Z$ boson%
\footnote{The results presented in this note are insensitive to
  non-zero width effects of the Higgs boson~\cite{nzw}.  For SM Higgs
  masses below $100$ GeV, $\Gamma_H$ is at least three orders of
  magnitude smaller than $\Gamma_Z$; for larger Higgs masses, $m_H$
  can be reinterpreted as the effective invariant mass of the Higgs
  decay products.}
is not taken into account, the cross section rises steeply at
threshold $\sim (s-(m_H+m_Z)^2)^{1/2}$.  After reaching a
maximum [about $25$ GeV above threshold in the LEP2 energy range], the
cross section falls off at high energies, according to the scaling law
$\sim g_W^4/s$ asymptotically.  Thus, Higgs-strahlung is the dominant
production process for moderate values of the energy.  The cross
section~(\ref{eq:HZ}) for Higgs-strahlung is reduced by a factor
$3\times {\rm BR}_\nu = 20\%$ if the final state of $Z$ decays is
restricted to neutrino pairs.

The total cross section for the $WW$ fusion of Higgs particles can be
cast into a similarly compact form~\cite{LC500}:
\begin{equation}\label{WWF}
  \sigma(e^{+}e^{-}\to \bar\nu_e\nu_e H)
  = \frac{G_{F}^{3}m_{W}^{4}}{4\sqrt{2}\,\pi^{3}}\,\int_{x_{H}}^{1}\,dx\,
    \int_{x}^{1}\,\frac{dy\,F(x,y)}{[1+(y-x)/x_{W}]^{2}}
\end{equation}
\begin{displaymath}
  F(x,y)
  = \left( \frac{2x}{y^{3}}-\frac{1+3x}{y^{2}}+\frac{2+x}{y}-1\right)\,
    \left[\frac{z}{1+z}-\log(1+z)\right]\,
    +\,\frac{x}{y^{3}}\frac{z^{2}(1-y)}{1+z}\nonumber
\end{displaymath}
where $x_H=m_H^2/s$, $x_W=m_W^2/s$ and $z=y(x-x_H)/(xx_W)$.  For
moderate Higgs masses and energies, the cross section, being ${\cal
O}(g_W^6)$, is suppressed with respect to Higgs-strahlung by the
additional electroweak coupling.  At high energies, the $WW$ fusion
process becomes leading, nevertheless, since the size of the cross
section is determined by the $W$ mass, in contrast to the
scale-invariant Higgs-strahlung process,
\begin{eqnarray}
  \sigma(e^{+}e^{-}\to \bar\nu_e\nu_e H)
  &\approx&  \frac{G_{F}^{3}m_{W}^{4}}{4\sqrt{2}\,\pi^{3}}
     \left[\left(1+\frac{m_H^2}{s}\right) \log \frac{s}{m_H^2}
     - 2\left(1-\frac{m_H^2}{s}\right)\right] \nonumber\\[2mm]
  &\to& \frac{G_{F}^{3}m_{W}^{4}}{4\sqrt{2}\,\pi^{3}}
     \log\frac{s}{m_H^2}
\end{eqnarray}
The cross section rises logarithmically at high energies, as to be
anticipated for this $t$-channel exchange process.

\begin{figure}[p]
\begin{center}
  \includegraphics{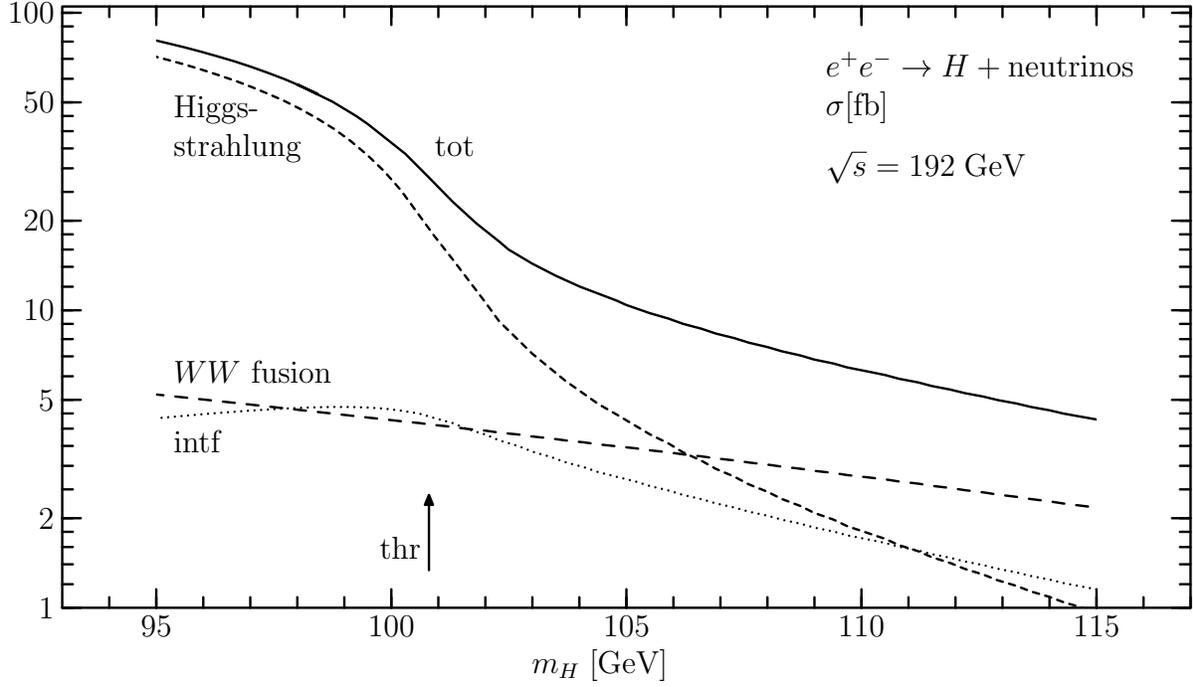}\\[15mm]
  \includegraphics{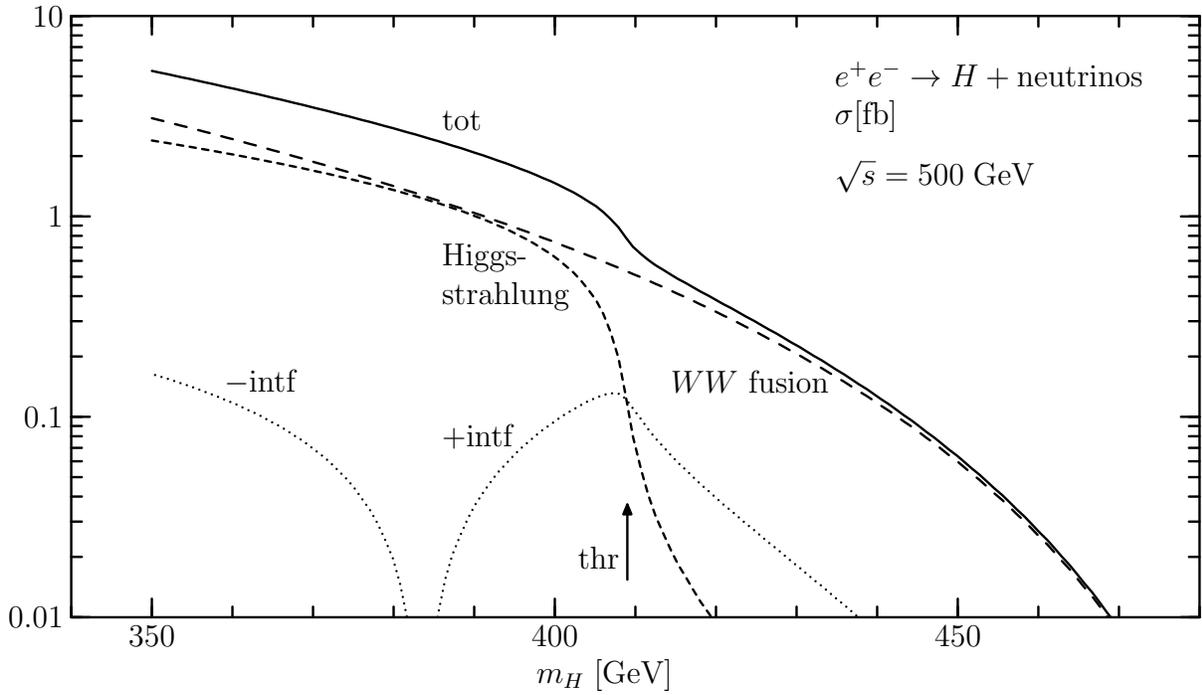}\\[8mm]
\end{center}
\caption{Total cross sections for the process $e^+e^-\to H\bar\nu\nu$
as a function of the Higgs mass.  The cross sections are broken down to
the three components Higgs-strahlung, $WW$ fusion, and the
interference term.  \emph{``thr''} denotes the maximum Higgs mass for
on-shell $ZH$ production, \emph{``tot''} is the total cross section.
For small Higgs masses the interference term is negative, for large
Higgs masses positive.}
\label{fig:sigma192/500}
\end{figure}

The compact form~(\ref{WWF}) for the fusion cross section has been
derived using the elegant method of invariant tensor
integration~\cite{cti}.  This method however cannot be applied any
more to calculate the interference term between $WW$ fusion and
Higgs-strahlung.  Nevertheless, similarly compact expressions can be
derived in this general case by choosing the energy $E_H$ and the
polar angle $\theta$ of the Higgs particle as the basic variables in
the $e^+e^-$ c.m.\ frame.  The overall cross section that will be
observed experimentally for the process
\begin{displaymath}
  e^+e^- \to H + \bar\nu\nu
\end{displaymath}
receives contributions $3\times{\cal G}_S$ from Higgs-strahlung with
$Z$ decays into three types of neutrinos, ${\cal G}_W$ from $WW$
fusion, and ${\cal G}_I$ from the interference term between fusion and
Higgs-strahlung for $\bar\nu_e\nu_e$ final states.  We find%
\footnote{The analytic result for ${\cal G}_W$ had first been
  obtained in Ref.\cite{AMP}.}
for energies $\sqrt{s}$ above the $Z$ resonance:
\begin{equation}\label{totalx}
  \frac{d\sigma(H\bar\nu\nu)}{dE_H\,d\cos\theta}
  = \frac{G_F^3 m_Z^8p}{\sqrt2\,\pi^3s}
  \left(3\,{\cal G}_S + {\cal G}_I + {\cal G}_W \right)
\end{equation}
with
\begin{eqnarray}
  {\cal G}_S &=& \frac{v_e^2+a_e^2}{96}\;
    \frac{ss_\nu + s_1s_2}{\left(s-m_Z^2\right)^2
               \left[(s_\nu-m_Z^2)^2 + m_Z^2\Gamma_Z^2\right]}\\
  {\cal G}_I &=& \frac{(v_e+a_e)\cos^4\theta_W}{8}\;
    \frac{s_\nu-m_Z^2}{\left(s-m_Z^2\right)
                \left[(s_\nu-m_Z^2)^2 + m_Z^2\Gamma_Z^2\right]}
    \nonumber\\
    &&\times
    \left[ 2 - (h_1+1)\log\frac{h_1+1}{h_1-1}
             - (h_2+1)\log\frac{h_2+1}{h_2-1}
           +\, (h_1+1)(h_2+1)\frac{{\cal L}}{\sqrt{r}}\right]
    \\
  {\cal G}_W &=& \frac{\cos^8\theta_W}{s_1 s_2 r}\,
    \Bigg\{(h_1+1)(h_2+1)
        \left[
        \frac{2}{h_1^2-1} + \frac{2}{h_2^2-1} - \frac{6s_\chi^2}{r}
        + \left(\frac{3t_1t_2}{r}-c_\chi\right)
          \frac{{\cal L}}{\sqrt{r}}\right]
    \nonumber\\
    &&\qquad\qquad
    - \left[\frac{2t_1}{h_2-1} + \frac{2t_2}{h_1-1}
      + \left(t_1+t_2+s_\chi^2\right)
        \frac{{\cal L}}{\sqrt{r}}\right]
    \Bigg\}
\end{eqnarray}
The cross section is written explicitly in terms of the Higgs momentum
$p= \sqrt{E_H^2-m_H^2}$, and the energy $\epsilon_\nu=\sqrt{s}-E_H$
and invariant mass squared $s_\nu=\epsilon_\nu^2-p^2$ of the neutrino
pair.  In addition, the following abbreviations have been adopted from
Ref.\cite{AMP},
\begin{displaymath}
  \begin{array}{r@{\;=\;}l}
  s_{1,2} & \sqrt{s}(\epsilon_\nu\pm p\cos\theta)\\[1mm]
  h_{1,2} & 1 + 2m_W^2/s_{1,2}\\[1mm]
  c_\chi & 1 - 2 s s_\nu/(s_1s_2)\\[1mm]
  s_\chi^2 & 1 - c_\chi^2
  \end{array}
  \qquad
  \begin{array}{r@{\;=\;}l}
  t_{1,2} & h_{1,2} + c_\chi h_{2,1}\\[1mm]
  r & h_1^2 + h_2^2 + 2c_\chi h_1h_2 - s_\chi^2\\[1mm]
  {\cal L} & {\displaystyle \log\frac{h_1h_2 + c_\chi + \sqrt{r}}
                        {h_1h_2 + c_\chi - \sqrt{r}}}
  \end{array}
\end{displaymath}

To derive the total cross section $\sigma(e^+e^-\to H\bar\nu\nu)$, the
differential cross section must be integrated over the region
\begin{equation}
  -1<\cos\theta<1 \quad\mbox{and}\quad
  m_H < E_H < \frac{\sqrt{s}}{2}\left(1+\frac{m_H^2}{s}\right)
\end{equation}
Since the compactification of the cross section requires tedious
analytical calculations, we have carefully cross-checked the result
for the total cross section by integrating numerically the squared
amplitude, computed by means of COMPHEP~\cite{COMPHEP}; the procedures
agreed at a level of $10^{-4}$ at all the points checked out.

\paragraph{}
To interpret the results, we display the three
components of the total cross section $\sigma(e^+e^-\to H\bar\nu\nu)$
in Fig.\ref{fig:sigma192/500} for the LEP2 energy $\sqrt{s}=192$
GeV and for the linear collider energy $\sqrt{s}=500$ GeV in the
cross-over regions.%
\footnote{Note that Higgs-strahlung dominates $WW$ fusion at $500$ GeV
  for moderate Higgs masses only if the total $ZH$ cross section is
  considered.}
It is obvious from the figures that
Higgs-strahlung, $WW$ fusion, and the interference term are of
comparable size in this region.

\begin{figure}[p]
\unitlength 1cm
\begin{center}
\begin{picture}(7.5,7.5)
\put(0.5,0){\includegraphics{e192.1}}
\end{picture}\hspace{1cm}
\begin{picture}(7.5,7.5)
\put(0.9,0){\includegraphics{e500.1}}
\end{picture}\\[0.8cm]
\end{center}
\vspace*{-.2\baselineskip}
\caption{Energy distribution of the Higgs bosons for the three
  components of the cross section [\/{\rm Hs} = Higgs-strahlung; $WW$
  = fusion;\/ {\rm intf} = interference term].  The individual curves
  are normalized to the total cross sections.  The\/ {\rm Hs} peak
  extends up to maximal values of\/ $0.52\,(0.22)\;{\rm GeV}^{-1}$\/
  for $\protect\sqrt{s} = 192$ and\/ $500\;{\rm GeV}$, respectively.
  The total cross sections are\/ $110.0\,(69.4)\;{\rm fb}$.}
\vspace*{-.6\baselineskip}
\label{fig:e}
\vspace{\baselineskip}
\begin{center}
\begin{picture}(7.5,7.5)
\put(0.5,0){\includegraphics{ct192.1}}
\end{picture}\hspace{1cm}
\begin{picture}(7.5,7.5)
\put(0.9,0){\includegraphics{ct500.1}}
\end{picture}\\[0.8cm]
\vspace*{-.2\baselineskip}
\caption{Angular distribution of the Higgs bosons [legend as
  Fig.\ref{fig:e}].}
\label{fig:ct}
\end{center}
\end{figure}

While the energy distribution of the Higgs particle peaks at $E_H\sim
(s+m_H^2-m_Z^2)/2\sqrt{s}$ for Higgs-strahlung, it is nearly flat for
$WW$ fusion~(Fig.\ref{fig:e}).  Only with rising total energy the
lower part of the Higgs spectrum becomes more pronounced.  The angular
distribution for Higgs-strahlung is almost isotropic at threshold
while the standard $\sin^2\theta$ law is approached, in accordance
with the equivalence principle, at asymptotic
energies~(Fig.\ref{fig:ct}).  The angular distribution peaks, by
contrast, in the $WW$ fusion process at $\theta\to 0$ and $\pi$ for
high energies as expected for $t$-channel exchange processes.

At linear colliders the incoming electron and positron beams can be
polarized longitudinally.  Higgs-strahlung and $WW$ fusion both
require opposite helicities of the electrons and positrons.  If
$\sigma_{U,L,R}$ denote the cross sections for unpolarized
electrons/positrons, left-handed electrons/right-handed positrons, and
right-handed electrons/left-handed positrons, respectively, we can
easily derive, in the notation of Eq.(\ref{totalx}):
\begin{eqnarray}
  \sigma_U &\propto&
     3\,{\cal G}_S + {\cal G}_I + {\cal G}_W\\
  \sigma_L &\propto&
     6\,{\cal G}_S + 4\,{\cal G}_I + 4\,{\cal G}_W\\
  \sigma_R &\propto&
     6\,{\cal G}_S
\end{eqnarray}
The cross section for $WW$ fusion of Higgs particles increases by a
factor four, compared with unpolarized beams, if left-handed electrons
and right-handed positrons are used.  By using right-handed electrons,
the $WW$ fusion mechanism is switched off.  [The interference term
cannot be separated from the $WW$ fusion cross section.]

It is trivial to transfer all these results from the Standard
Model to the Minimal Supersymmetric Standard Model (MSSM).  Since the
couplings to $Z/W$ gauge bosons in the MSSM are shared~\cite{GH} by
the CP-even light and heavy scalar Higgs bosons, $h$ and $H$,
respectively, only the overall normalization of the cross sections is
modified with respect to the Standard Model:
\begin{eqnarray}
  \sigma(h)_{\rm MSSM} &=& \sin^2(\beta-\alpha)\times\sigma(H)_{\rm SM} \\
  \sigma(H)_{\rm MSSM} &=& \cos^2(\beta-\alpha)\times\sigma(H)_{\rm SM}
\end{eqnarray}
Higgs-strahlung, $WW$ fusion, and the interference term are affected in
the same way.  [The angle $\alpha$ is the mixing angle in the CP-even
Higgs sector while the mixing angle $\beta$ is determined by the
ratio of the vacuum expectation values of the two neutral Higgs
fields in the MSSM.  A recent discussion of the size of the
coefficients $\sin^2/\cos^2(\beta-\alpha)$ may be
found in Ref.\cite{DKZ}.]

\subsection*{Acknowledgement}
We are very grateful to V.A.\ Khoze for the critical reading of the
manuscript and, in particular, for emphasizing the usefulness of
polarized beams in Higgs production.


\end{document}